\documentclass[british,english,american,12pt]{iopart}
\usepackage[latin9]{inputenc}
\usepackage{geometry}
\usepackage{color}
\usepackage{babel}
\usepackage{prettyref}
\usepackage{amstext}
\usepackage{amssymb}
\usepackage{graphicx}
\usepackage[unicode=true,pdfusetitle,
 bookmarks=true,bookmarksnumbered=false,bookmarksopen=false,
 breaklinks=false,pdfborder={0 0 1},backref=false,colorlinks=true]
 {hyperref}
\hypersetup{
 colorlinks=true,linkcolor=blue,citecolor=blue,urlcolor=blue,filecolor=blue}
\usepackage{breakurl}

\makeatletter

\newcommand{\lyxdot}{.}

\usepackage{iopams}
\usepackage{setstack}

\usepackage[numbers, sort&compress]{natbib}
\usepackage{iopams}

\def\newblock{\hskip .11em plus .33em minus .07em}

\newrefformat{cap}{\hyperref[#1]{Figure~\ref{#1}}}
\newrefformat{fig}{\hyperref[#1]{Figure~\ref{#1}}}
\newrefformat{tab}{\hyperref[#1]{Table~\ref{#1}}}
\newrefformat{sec}{\hyperref[#1]{Sec.~\ref{#1}}}
\newrefformat{sub}{\hyperref[#1]{Sec.~\ref{#1}}}
\newrefformat{alg}{\hyperref[#1]{Alg.~\ref{#1}}}
\newrefformat{app}{\hyperref[#1]{\ref{#1}}}

\newcommand{\eqref}[1]{\eref{#1}}

\makeatother

\begin{document}
\global\long\def\c{\mathbf{c}}
\global\long\def\b{\mathbf{b}}
\global\long\def\B{\mathbf{B}}
\global\long\def\F{\mathbf{F}}
\global\long\def\n{\mathbf{n}}
\global\long\def\P{\mathbf{P}}
\global\long\def\C{\mathbf{C}}
\global\long\def\M{\mathbf{M}}
\global\long\def\D{\mathbf{D}}
\global\long\def\W{\mathbf{W}}
\global\long\def\Q{\mathbf{Q}}
\global\long\def\I{\mathbf{1}}
\global\long\def\r{\mathbf{r}}
\global\long\def\a{\mathbf{a}}
\global\long\def\h{\mathbf{h}}
\global\long\def\e{\mathbf{e}}
\global\long\def\s{\mathbf{s}}
\global\long\def\Ew{\mathrm{E}}
\global\long\def\Ex{\mathrm{e}}
\global\long\def\In{\mathrm{i}}
\global\long\def\Exc{\mathcal{E}}
\global\long\def\Inh{\mathcal{I}}
\global\long\def\taue{\tau_{e}}
\global\long\def\ext{\text{ext.}}
\global\long\def\comm{\text{comm}}
\global\long\def\taua{\tau_{a}}
\global\long\def\Fourier{\mathcal{F}}
\global\long\def\defeq{\stackrel{\mathrm{def}}{=}}
\global\long\def\mV{\;\mathrm{mV}}
\global\long\def\ms{\;\mathrm{ms}}
\global\long\def\sec{\;\mathrm{s}}
\global\long\def\Hz{\;\mathrm{Hz}}
\global\long\def\taur{\tau_{r}}
\global\long\def\taum{\tau_{m}}
\global\long\def\taus{\tau_{s}}
\global\long\def\Rm{R_{m}}
\global\long\def\Cm{C_{m}}

\title[Echoes in correlated neural systems]{Echoes in correlated neural systems}

\author{{\Large M Helias$^{1}$, T Tetzlaff$^{1}$ and M Diesmann$^{1,2,3}$}}

\address{{\Large $^{1}$}{\large Institute of Neuroscience and Medicine (INM-6)
and Institute for Advanced Simulation (IAS-6)}\\
{\large J{\"u}lich Research Centre and JARA, J{\"u}lich, Germany}}

\address{{\Large $^{2}$}{\large RIKEN Brain Science Institute, Wako, Saitama,
Japan}}

\address{{\large $^{3}$Medical Faculty, RWTH Aachen University, Germany}}

\ead{{\large m.helias@fz-juelich.de}}
\begin{abstract}
Correlations are employed in modern physics to explain microscopic
and macroscopic phenomena, like the fractional quantum Hall effect
and the Mott insulator state in high temperature superconductors and
ultracold atoms. Simultaneously probed neurons in the intact brain
reveal correlations between their activity, an important measure to
study information processing in the brain that also influences macroscopic
signals of neural activity, like the electro encephalogram (EEG).
Networks of spiking neurons differ from most physical systems: The
interaction between elements is directed, time delayed, mediated by
short pulses, and each neuron receives events from thousands of neurons.
Even the stationary state of the network cannot be described by equilibrium
statistical mechanics. Here we develop a quantitative theory of pairwise
correlations in finite sized random networks of spiking neurons. We
derive explicit analytic expressions for the population averaged cross
correlation functions. Our theory explains why the intuitive mean
field description fails, how the echo of single action potentials
causes an apparent lag of inhibition with respect to excitation, and
how the size of the network can be scaled while maintaining its dynamical
state. Finally, we derive a new criterion for the emergence of collective
oscillations from the spectrum of the time-evolution propagator.
\end{abstract}




\submitto{\NJP }

\maketitle

\section{Introduction}

Correlations are an established feature of neural activity \cite{Gerstein69_828}
and evidence increases that they can provide insights into the information
processing in the brain \cite{Cohen11_811}. The temporal relationship
between the activity of pairs of neurons is described by correlation
functions. Their shape has early been related to the direct coupling
between neurons and to the common input shared by pairs of neurons.
On the one hand, correlations may limit the signal-to-noise ratio
of population rate signals \cite{Zohary94_140}, on the other hand
they have been shown to increase the amount of information available
to unbiased observers \cite{Abbott99_91}. Furthermore, synchronous
neural activity has been proposed to bind elementary representations
into more complex objects \cite{Malsburg86} and experimental evidence
for such a correlation code is provided by task related modulation
of synchrony in primary visual cortex \cite{Maldonado08_1523} and
in motor cortex \cite{Kilavik09_12653}.

The small magnitude \cite{Ecker10} of pairwise correlations in the
asynchronous irregular state \cite{Brunel00_183} of cortex has recently
been related to the balance between excitation and inhibition in local
networks \cite{Hertz10_427,Renart10_587} and inhibitory feedback
was identified as a general mechanism of decorrelation \cite{Tetzlaff12_e1002596}.
However, a quantitative theory explaining the temporal shape of correlation
functions in recurrently impulse coupled networks of excitatory and
inhibitory cells remained elusive.

\begin{figure}
\selectlanguage{british}%
\begin{raggedleft}
\includegraphics[scale=0.8]{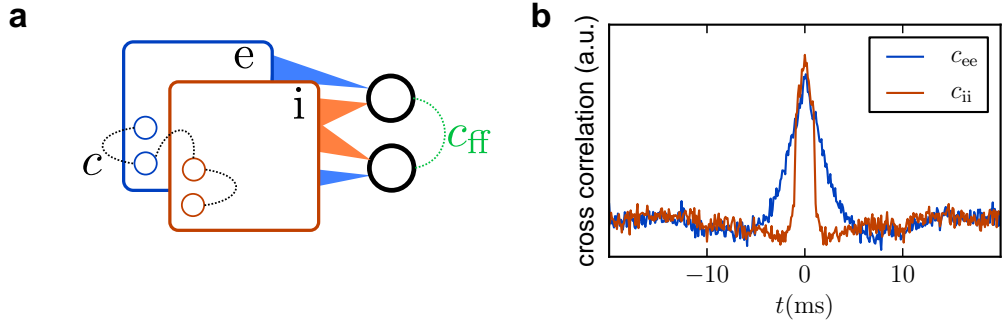}
\par\end{raggedleft}

\selectlanguage{american}%
\begin{centering}

\par\end{centering}

\caption{Self-consistency argument fails for covariances in a homogeneous recurrent
random network. \textbf{(a)} Each neuron (black circles) receives
input from the same number of randomly chosen excitatory ($\mathrm{e}$)
and inhibitory ($\mathrm{i}$) neurons in the network, so the input
statistics of all neurons is the same. The covariance $c$ within
the network determines the covariance between the inputs to a pair
of neurons and hence the covariance $c_{\mathrm{ff}}$ of their outputs.
Self-consistency seems to require $c_{\mathrm{ff}}=c=c_{\mathrm{ee}}=c_{\mathrm{ii}}$.
\textbf{(b)} Covariance functions averaged over pairs of excitatory
($c_{\mathrm{ee}}$) and over pairs of inhibitory ($c_{\mathrm{ii}}$)
integrate-and-fire model neurons are different in a direct simulation.
Other parameters are given in \prettyref{app:simulation_parameters}.\foreignlanguage{british}{\label{fig:correlation_transmission}}}
\end{figure}

Assuming random connectivity with identical numbers and strengths
of incoming synapses per neuron, as illustrated in \prettyref{fig:correlation_transmission},
suggests by mean field arguments \cite{Vreeswijk96,Amit97} that the
resulting activity of two arbitrarily selected neurons and hence the
power spectra of activities averaged over excitatory or inhibitory
neurons should be the same. Direct simulations, however, exhibit different
power spectra for these sub-populations \cite{Kriener08_2185}. A
similar argument holds for the covariance $c_{\mathrm{ff}}$ between
the two neurons: If the covariance $c$ between any pair of inputs
is known, the covariance between their outgoing activity $c_{\mathrm{ff}}$
is fully determined \cite{Shadlen98,Stroeve01,Tetzlaff02,Morenobote06_028101,DeLaRocha07_802,Shea-Brown08,Kriener08_2185,Burak09_2269,Rosenbaum10_00116,Tchumatchenko10_058102}.
By self-consistency, as both neurons belong to the same recurrent
network, one concludes that $c_{\mathrm{ff}}=c$. In particular the
covariance averaged over excitatory pairs should be identical to the
corresponding average over inhibitory pairs, which is in contrast
to direct simulation (\prettyref{fig:correlation_transmission}b).
In this work, we elucidate why this mean field argument for covariances
fails and derive a self-consistency equation for pairwise covariances
in recurrent random networks which explains the differences in the
power spectra and covariances.

Theories for pairwise covariances have been derived for binary neuron
models \cite{Ginzburg94,Renart10_587} and for excitatory stochastic
point process models \cite{Hawkes71_438}. However, the lack of either
inhibition or delayed pulsed interaction limits the explanatory power
of these models. A theory for networks of leaky integrate-and-fire
(LIF) model neurons \cite{Stein65} is required, because this model
has been shown to well approximate the properties of mammalian pyramidal
neurons \cite{Rauch03} and novel experimental techniques allow to
reliably assess the temporal structure of correlations in cortex \cite{Okun_535_08}.
Moreover, the relative timing of action potentials is the basis for
models of synaptic plasticity \cite{Morrison08_459}, underlying learning
in biological neural networks. Analytical methods to treat population
fluctuations in spiking networks are well advanced \cite{Cai12_307}
and efficient hybrid analytical-numerical schemes exist to describe
pairwise covariances \cite{Cai2004_14288}. Here we present an analytically
solvable theory of pairwise covariances in random networks of spiking
leaky integrate-and-fire model neurons with delayed pulsed interaction
in the asynchronous irregular regime.

\section{Results}

We consider recurrent random networks of $N$ excitatory and $\gamma N$
inhibitory leaky integrate-and-fire model neurons receiving pulsed
input (spikes) from other neurons in the network. Each neuron has
$K=pN$ incoming excitatory synapses independently and randomly drawn
from the pool of excitatory neurons, and $\gamma K=\gamma pN$ inhibitory
synapses (homogeneous Erd\H{o}s-R\'{e}nyi random network with fixed
in-degree). An impulse at time $t$ arrives at the target neuron after
the synaptic delay $d$ and elicits a synaptic current $I_{i}$ that
decays with time constant $\taus$ and causes a response in the membrane
potential $V_{i}$ (with time constant $\taum)$ proportional to the
synaptic efficacy $J$ (excitatory) or $-gJ$ (inhibitory), respectively.
The coupled set of differential equations governing the subthreshold
dynamics of a single neuron $i$ is \cite{Fourcaud02}
\begin{eqnarray}
\taum\frac{dV_{i}}{dt} & = & -V_{i}+I_{i}(t)\nonumber \\
\taus\frac{dI_{i}}{dt} & = & -I_{i}+\taum\sum_{j=1,j}^{N}J_{ij}s_{j}(t-d),\label{eq:diffeq_iaf}
\end{eqnarray}
where the membrane resistance was absorbed into $J_{ij}$. If $V_{i}$
reaches the threshold $V_{\theta}$ at time point $t_{k}^{i}$ the
neuron emits an action potential and the membrane potential is reset
to $V_{r}$, where it is clamped for the refractory time $\taur$.
The spiking activity of neuron $i$ is described by this sequence
of action potentials, the spike train $s_{i}(t)=\sum_{k}\delta(t-t_{k}^{i})$.

\begin{figure}
\begin{raggedleft}
\includegraphics[scale=0.8]{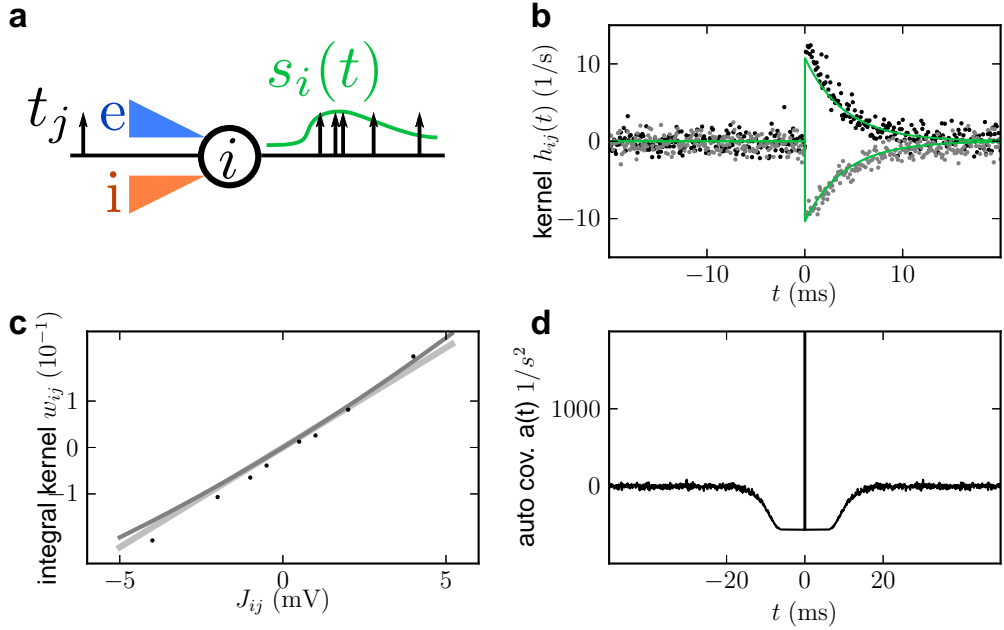}
\par\end{raggedleft}

\caption{\selectlanguage{american}%
Mapping the integrate-and-fire dynamics to a linear coupling kernel.
\textbf{(a)} The kernel $h_{ij}$ determines the transient effect
of an incoming impulse at time point $t_{j}$ (black arrow) on the
density $s_{i}(t)$ of outgoing action potentials, averaged over realizations
of the stochastic activity of the remaining inputs (indicated as red
and blue triangles). \textbf{(b)} Response kernel \eqref{eq:imp_response}
(green) compared to direct simulation for an impulse of amplitude
$J=1\mV$ (black dots) and $J=-1\mV$ (gray dots). Time constant $\taue=4.07\ms$
determined by a least squares fit to a single exponential. The background
activity causes a mean $\mu_{i}=15\mV$ and fluctuations $\sigma_{i}=10\mV$.\textbf{
(c)} Linear and quadratic dependence \prettyref{eq:w_ij} of the integral
response $w_{ij}$ on $J_{ij}$ (dark gray curve) and linear term
alone (light gray line). \textbf{(d)} Autocovariance function of the
spike train with a $\delta$ peak at $t=0$ and covariance trough
due to refractoriness.\foreignlanguage{english}{\label{fig:response_kernel}}\selectlanguage{british}%
}
\end{figure}

The activity of a given neuron $i$ depends on the history of the
other neurons' activities $\s(t)=(s_{1}(t),\ldots,s_{N}(t))^{T}$
in the network, so formally we can consider the spike train of neuron
$i$ as a functional of all other spike trains. The time averaged
covariance matrix expresses these interrelations and is defined as
$\bar{\c}(\tau)=\left\langle \s(t+\tau)\s^{T}(t)\right\rangle _{t}-\r\r^{T}$,
where $\text{ }\r=\langle\s\rangle_{t}$ is the vector of time averaged
firing rates. The diagonal contains the autocovariance functions (diagonal
matrix $\a(\tau)$) which are dominated by a $\delta$-peak at zero
time lag and for $\tau\neq0$ exhibit a continuous shape mostly determined
by refractoriness, the inability of the neuron to fire spikes in short
succession due to the voltage reset, as shown in \prettyref{fig:response_kernel}d.
The off-diagonal elements contain the cross covariance functions $\c(\tau)$
that originate from interactions. We therefore decompose the covariance
matrix into $\bar{\c}=\a+\c$. A basic property of covariance matrices
is the symmetry $\bar{\c}(\tau)=\bar{\c}^{T}(-\tau)$, so we only
need to consider $\tau>0$ and obtain the solution for $\tau<0$ by
symmetry. Each spike at time $t^{\prime}$ can influence the other
neurons at time $t>t^{\prime}$. Formally we express this influence
of the history up to time $t$ on a particular neuron $i$ in the
network as $s_{i}(t,\{\s(t^{\prime})|t^{\prime}<t\})$, which is a
functional of all spike times until $t$. In the asynchronous state
of the network \cite{Brunel00_183} and for small synaptic amplitudes
$J_{ij}$ a single spike of neuron $j$ causes only a small perturbation
of $s_{i}$. The other inputs to neuron $i$ effectively act as additional
noise. We may therefore perform a linear approximation of $j$'s direct
influence on neuron $i$ and average over the realizations of the
remaining inputs $\s\backslash s_{j}$, as illustrated in \prettyref{fig:response_kernel}a.
This linearization and the superposition principle lead to the convolution
equation
\begin{eqnarray}
\langle s_{i}(t)|s_{j}\rangle_{\mathrm{\s\backslash s_{j}}} & = & r_{i}+\int_{-\infty}^{t}h_{ij}(t,t^{\prime})(s_{j}(t^{\prime})-r_{j})\, dt^{\prime}\label{eq:lin_approx_F}\\
 & = & r_{i}+[h_{ij}\ast(s_{j}-r_{j})](t),\nonumber 
\end{eqnarray}
where we define as the linear response kernel $h_{ij}(t,t^{\prime})=\left\langle \frac{\delta s_{i}(t)}{\delta s_{j}(t^{\prime})}\right\rangle _{\s\backslash s_{j}}$
the functional derivative of $s_{i}(t)$ with respect to $s_{j}(t^{\prime})$,
formally defined in \prettyref{app:response_kernel}. The kernel $h_{ij}$
quantifies the effect of a single spike at time $t^{\prime}$ of neuron
$j$ on the expected density of spikes of neuron $i$ at time $t$
by a direct synaptic connection from neuron $j$ to neuron $i$, earlier
introduced as the ``synaptic transmission curve'' \cite{Abeles91}.
This density vanishes for $t<t^{\prime}$ due to causality. For the
stationary network state studied here, the kernel further only depends
on the time difference $\tau=t-t^{\prime}$. In a consistent linear
approximation there are no higher order terms, so the effects of two
inputs $s_{j}$ and $s_{k}$ superimpose linearly. In general, the
response of a cell is typically supra-linear in the number of synchronously
arriving excitatory spikes \cite{Abeles82b,Abeles91,Goedeke08_015007}.
If, however, the network state to be described is sufficiently asynchronous,
as is the case here, a linear approximation is adequate. Using this
linear expansion and the definition of the covariance matrix, the
off-diagonal elements of the covariance matrix fulfill a linear convolution
equation 
\begin{eqnarray}
\c(\tau) & = & [\h\ast\left(\a+\c\right)](\tau)\quad\text{for }\tau>0.\label{eq:integral_equation_corrfunction}
\end{eqnarray}
The equation is by construction valid for $\tau>0$ and needs to be
solved simultaneously obeying the symmetry condition $\c(t)=\c(-t)^{T}$.
Up to linear order in the interaction, equation \eqref{eq:integral_equation_corrfunction}
is the correct replacement of the intuitive self-consistency argument
sketched in \prettyref{fig:correlation_transmission}.

In order to relate the kernel $h_{ij}$ to the leaky integrate-and-fire
model, we employ earlier results based on Fokker-Planck theory \cite{Fourcaud02}.
For small synaptic amplitudes $J\ll V_{\theta}-V_{r}$ and weak pairwise
covariances the summed synaptic input $\taum\sum_{j=1}^{N}J_{ij}s_{j}(t-d)$
can be approximated as a Gaussian white noise with mean $\mu_{i}=\taum\sum_{j}J_{ij}r_{j}$
and variance $\sigma_{i}^{2}=\taum\sum_{j}J_{ij}^{2}r_{j}$ for neurons
firing with rates $r_{j}$ and Poisson statistics. For short synaptic
time constants $\taus\ll\taum$ the stationary firing rate $r_{i}(\mu_{i},\sigma_{i}^{2})$
\prettyref{eq:rate} depends on these two moments \cite{Fourcaud02}.
In a homogeneous random recurrent network the input to each neuron
is statistically the same, so the stationary rate $r_{i}=r$ is identical
for all neurons. It is determined by the self-consistent solution
of $r(\mu_{i},\sigma_{i}^{2})$, taking into account the dependence
of $\mu_{i}$ and $\sigma_{i}^{2}$ on the rate $r$ itself \cite{Brunel00_183}. 

The integral $w_{ij}=\int_{0}^{\infty}h_{ij}(t)\, dt$ of the response
kernel is equivalent to the DC susceptibility $w_{ij}=\frac{\partial r_{i}}{\partial r_{j}}$
\cite{Helias10_1000929} and has earlier been termed ``asynchronous
gain'' \cite{Abeles91}. The approximation is second order in the
synaptic amplitude $w_{ij}=\alpha J_{ij}+\beta J_{ij}^{2}$. The first
order term originates from the dependence of $\mu_{i}$ on $r_{j}$,
the second order term stems from the dependence of $\sigma_{i}^{2}$
on $r_{j}$ \eqref{eq:w_ij}.

\selectlanguage{british}%

\selectlanguage{american}%
Throughout this work we choose the working point of the neurons in
the network such that the firing of the cells is driven by strong
fluctuations and the mean membrane potential is close to threshold,
in order to have irregular firing. This is achieved by appropriate
choices of the external driving Poisson sources, as described in \prettyref{app:simulation_parameters}.
For the small networks considered here, it is realistic to assume
that about $50$ percent of the inputs to a neuron come from outside
the local network \cite{Stepanyants09_3555}. \prettyref{fig:response_kernel}b
shows the deflection of the firing rate from baseline caused by an
impulse in the input averaged over many repetitions. For sufficiently
strong fluctuations $\sigma_{i}$, the quadratic term in $J_{ij}$
is negligible, as seen from \prettyref{fig:response_kernel}c. For
smaller membrane potential fluctuations $\sigma_{i}$ we expect the
linear approximation to be less accurate.

The kernel shows exponential relaxation with an effective time constant
$\taue$ that depends on the working point $(\mu_{i},\sigma_{i})$
and the parameters of the neuron, which is obtained by a least squares
fit of a single exponential to the simulated response in \prettyref{fig:response_kernel}b
for one particular amplitude $J_{ij}$. We therefore approximate the
response $h_{ij}(t)$ as
\begin{eqnarray}
w_{ij}h(t) & = & \Theta(t-d)\;\frac{w_{ij}}{\tau_{e}}e^{-\frac{t-d}{\tau_{e}}},\label{eq:imp_response}
\end{eqnarray}
where $d$ is the synaptic delay and $\Theta$ the Heaviside step
function.

In experiments covariance functions are typically averaged over statistically
equivalent pairs of neurons. Such averages are important, because
they determine the behavior of the network on the macroscopic scale
of populations of neurons. We therefore aim at a corresponding effective
theory. Assuming identical dynamics \eqref{eq:diffeq_iaf}, all neurons
have, to good approximation, the same autocovariance function $a(t)$
and response kernel $h(t)$. So each incoming excitatory impulse causes
a response $w\, h(t)$, an inhibitory impulse $-gw\, h(t)$. We define
the covariance function averaged over all pairs of excitatory neurons
as $c_{\Ex\Ex}(\tau)=\frac{1}{N^{2}}\sum_{i,j\in\Exc,i\neq j}c_{ij}(\tau),$
(setting $N(N-1)\simeq N^{2}$ for $N\gg1$), where $\Exc$ denotes
the set of all excitatory neurons. The pairings $c_{\Ex\In},c_{\In\Ex},$
and $c_{\In\In}$ are defined analogously. Inserting equation \eqref{eq:integral_equation_corrfunction}
into the average $c_{\Ex\Ex}(\tau)$, the first term proportional
to the autocovariance $a(t)$ only contributes if neuron $j$ projects
to neuron $i$. For fixed $i$, there are $K$ such indices $j$,
so the first term yields $\sum_{i,j\in\Exc,i\neq j}h_{ij}\ast a_{j}=NKw\, h\ast a$.
The second sum $\sum_{i,j\in\Exc,i\neq j}h_{ik}\ast c_{kj}$ can be
decomposed into a sum over all intermediate excitatory neurons $k\in\Exc$
and over all inhibitory neurons $k\in\Inh$ projecting to neuron $i$.
Replacing the individual covariances by their population average,
$c_{\Ex\Ex}$ and $c_{\In\Ex}$, respectively, and considering the
number of connections and their amplitude we obtain $wNK\, h\ast\left(c_{\Ex\Ex}-\gamma g\, c_{\In\Ex}\right)$,
with $\gamma=N_{\In}/N_{\Ex}$ and $g=w_{\In}/w_{\Ex}$ the relative
number of inhibitory neurons and the relative linearized inhibitory
synaptic amplitude. Similar relations hold for the remaining three
averages, so we arrive at a two-by-two convolution matrix equation
for the pairwise averaged covariances for $\tau>0$

\begin{eqnarray}
\c(\tau) & = & \left[h\ast\M\c\right](\tau)+\Q\left[h\ast a\right](\tau)\label{eq:integral_equation_corrfunction_avg}\\
\text{with }\M & = & Kw\left(\begin{array}{cc}
1 & -\gamma g\\
1 & -\gamma g
\end{array}\right),\quad\Q=\frac{Kw}{N}\left(\begin{array}{cc}
1 & -g\\
1 & -g
\end{array}\right),\nonumber \\
\text{and }\c(\tau) & = & \left(\begin{array}{cc}
c_{\Ex\Ex}(\tau) & c_{\Ex\In}(\tau)\\
c_{\In\Ex}(\tau) & c_{\In\In}(\tau)
\end{array}\right).\nonumber 
\end{eqnarray}

The convolution equation only holds for positive time lags $\tau$.
For negative time lags it is determined by the symmetry $\c(-\tau)=\c^{T}(\tau)$.
The solution of this equation can be obtained by an extension of the
method used in \cite{Hawkes71_438} employing Wiener-Hopf theory \cite{Hazewinkel02}
to the cross spectrum $\C(\omega)=\int_{-\infty}^{\infty}\c(t)e^{-i\omega t}\, dt$
in frequency domain, as shown in \prettyref{app:covariance_general_autocorrelation}
(here capital letters denote the Fourier transform of the respective
lower case letters). With the definition of the propagator $\P(\omega)=\left(\I-\M H(\omega)\right)^{-1}$
the cross spectrum takes the form $\C(\omega)=\P(\omega)\left[D_{+}(\omega)\Q+D_{+}(-\omega)\Q^{T}-A(\omega)|H(\omega)|^{2}\Q\M^{T}\right]\P^{T}(-\omega),$
where we split the term $H(\omega)\; A(\omega)=D_{+}(\omega)+D_{-}(\omega)$
so that $d_{+}(\tau)$ and $d_{-}(\tau)$ vanish for times $\tau<0$
and $\tau>0$, respectively. For the averaged cross spectrum, the
matrix $\Q$ is defined in \eqref{eq:integral_equation_corrfunction_avg},
the non-averaged cross spectrum can be recovered as a special case
setting $\Q=\M=\W$, because the convolution equations \eqref{eq:integral_equation_corrfunction}
and \eqref{eq:integral_equation_corrfunction_avg} have the same structure
and symmetries. If all eigenvalues of $\M H(\omega)$ have an absolute
value smaller than unity, the propagator $\P(\omega)$ can be expanded
into a geometric series in which the $n$-th term contains only interactions
via $n$ steps in the connectivity graph. This expansion has been
used to obtain the contribution of different motifs to the integral
of covariance functions \cite{Pernice11_e1002059} and their temporal
shape \cite{Trousdale12_e1002408}.

\selectlanguage{english}%

\selectlanguage{american}%

In the following, we neglect the continuous part of the autocovariance
function, setting $a(t)=r\;\delta(t)$, because the $\delta$-peak
is typically dominant. With this replacement we mainly neglect the
trough around $0$ due to the relative refractoriness. An estimate
of the error can be obtained considering the respective weights of
the delta peak ($r$) and the through. From the relation $\int_{-\infty}^{\infty}a(t)\, dt=r\mathrm{CV^{2}}$
\cite{Rieke97,MorenoBote08} the integral weight of the trough follows
as $r(\mathrm{CV}^{2}-1)$, which is small for irregular spike trains
with a coefficient of variation $\mathrm{CV}$ close to unity. 

\begin{figure}
\selectlanguage{british}%
\raggedleft{}\includegraphics[scale=0.8]{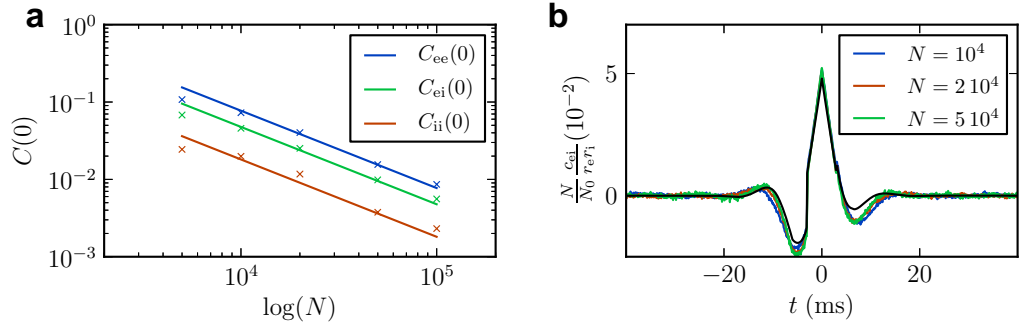}\foreignlanguage{american}{\caption{Shape invariance of covariances with network scale. Synapses scaled
as $J\propto1/N$ with network size $N$ to conserve the population
feedback $L=\mathrm{const}.$ \textbf{(a)} Integral correlation coefficient
averaged over different pairs of neurons and theory \eqref{eq:corrtrans_theory}
confirming the $\propto1/N$ dependence. \textbf{(b) }Rescaled covariance
functions $Nc_{\Ex i}$ averaged over excitatory-inhibitory pairs
of neurons for different network sizes (color coded) and theory \eqref{eq:covariance_time}
(black).\foreignlanguage{british}{ }\label{fig:scaling_corr_functions}}
}\selectlanguage{american}%
\end{figure}

For an arbitrary causal kernel $h$ it follows that $D_{+}(\omega)=r\, H(\omega),$
so the cross spectrum takes the form
\begin{eqnarray}
\C(\omega) & = & r\frac{Kw}{N}\left(\begin{array}{cc}
1 & -g\\
1 & -g
\end{array}\right)U(i\omega)+\text{c.c. trans. }\label{eq:coherence_causal_A}\\
 & + & r(1+g^{2}\gamma)\frac{(Kw)^{2}}{N}\left(\begin{array}{cc}
1 & 1\\
1 & 1
\end{array}\right)|U(i\omega)|^{2}\nonumber \\
\text{with } & U(z) & =\frac{1}{H^{-1}(-iz)-L}\nonumber \\
\text{and } & L & =Kw(1-\gamma g).\nonumber 
\end{eqnarray}
The limit $\omega\rightarrow0$ corresponds to the time integral of
the cross covariance function, approximating the count covariance
for long time bins \cite{Cohen11_811,Ecker10}. With $A(0)=r$, the
integral correlation coefficient averaged over neuron pairs fulfills
the equation

\begin{eqnarray}
\frac{\C(0)}{A(0)} & = & \frac{Kw}{N}\frac{1}{1-L}\left(\begin{array}{cc}
2 & 1-g\\
1-g & -2g
\end{array}\right)\label{eq:corrtrans_theory}\\
 &  & +\frac{(Kw)^{2}}{N}\frac{1+g^{2}\gamma}{(1-L)^{2}}\left(\begin{array}{cc}
1 & 1\\
1 & 1
\end{array}\right),\nonumber 
\end{eqnarray}
which has previously been derived from a noise-driven linear rate
dynamics \cite{Tetzlaff12_e1002596}. The quantity $L$ plays a key
role here: It determines the feedback magnitude of in-phase fluctuations
of the excitatory and inhibitory population. Stability of the average
firing rate requires this feedback to be sufficiently small \cite{Brunel00_183},
i.e. $L<1$, indicated by the pole in equation \eqref{eq:corrtrans_theory}
at $L=1$. Typically cortical networks are in the balanced regime
\cite{Vreeswijk96,Amit97,Brunel00_183}, i.e. $L<0$. For such inhibition
dominated networks, the denominator in equation \eqref{eq:corrtrans_theory}
is larger than unity, indicating a suppression of covariances \cite{Tetzlaff12_e1002596}.
As shown in \prettyref{fig:scaling_corr_functions}a, the prediction
\prettyref{eq:corrtrans_theory} agrees well with the results of simulations
of leaky integrate-and-fire networks for a broad range of network
sizes $N$.

Previous works have investigated neural networks in the thermodynamic
limit $N\rightarrow\infty$ \cite{Vreeswijk96,Renart10_587}, scaling
the synaptic amplitudes $J\propto1/\sqrt{N}$ in order to arrive at
analytical results. Such a scaling increases the feedback on the network
level $L\propto\sqrt{N}$ and therefore changes the collective network
state. Equation \eqref{eq:corrtrans_theory} provides an alternative
criterion to scale the synapses while keeping the dynamics comparable:
The two terms in equation \eqref{eq:corrtrans_theory} depend differently
on the feedback $L$. In order to maintain their ratio, we need to
keep the population feedback $L$ constant. The synaptic amplitude
$J$ (approximately $\propto w$ \prettyref{eq:w_ij}) hence needs
to scale as $J\propto1/N.$ In addition, the response kernel $h$
of each single neuron must remain unchanged, requiring the same working
point, characterized by the mean $\mu_{i}$ and fluctuations $\sigma_{i}$
in the input into each cell. Constant mean directly follows from $L=\mathrm{const.}$,
but the variance due to local input from other neurons in the network
decreases as $1/N$. To compensate, we supply each neuron with an
additional external uncorrelated balanced noise whose variance appropriately
increases with $N$ (as described in detail in \prettyref{app:simulation_parameters}).
\prettyref{fig:scaling_corr_functions}b shows that the shape of the
covariance functions is invariant over a large range of network sizes
$N$, in particular the apparent time lag of inhibition behind excitation
observed as the asymmetry in \prettyref{fig:scaling_corr_functions}b
does not vanish in the limit $N\rightarrow\infty$. The magnitude
of the covariance decreases as $1/N$ as expected from equation \eqref{eq:corrtrans_theory},
because $Kw=\mathrm{const.}$

\begin{figure}
\selectlanguage{british}%
\begin{raggedleft}
\includegraphics[scale=0.8]{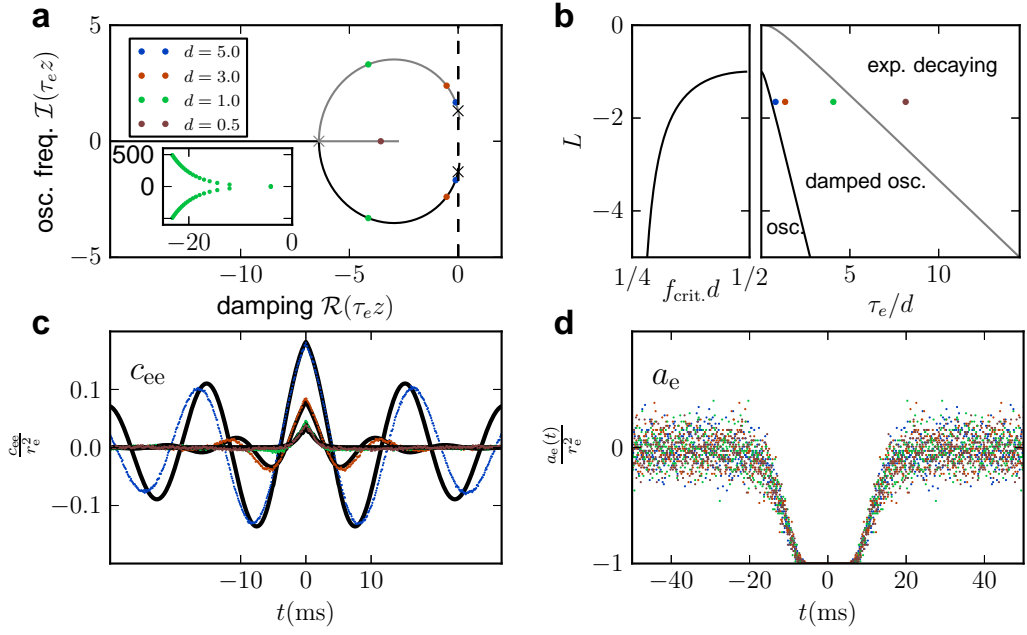}
\par\end{raggedleft}

\selectlanguage{american}%
\centering{}\caption{Phase diagram determined by spectral analysis. Throughout all panels
colors correspond to delays as given in \textbf{a}.\textbf{ (a)} Each
dot in the inset represents a pole $z_{k}$ \eqref{eq:pole_r} for
delay $d=1\ms$ (two rightmost poles appear as one point). The two
rightmost poles change with delay $d$. At $d=0.753\ms$ (gray St.
Andrew's cross) \eqref{eq:principal_splitting} the poles become a
conjugate pair, at $d=6.88\ms$ (black crosses) \eqref{eq:omega_d_crit}
both poles have a zero real part, causing oscillations (Hopf bifurcation).
\textbf{(b)} Right: Phase diagram spanned by $\taue/d$ and feedback
$L$. Onset of oscillations below the black curve \eqref{eq:omega_d_crit}
(Hopf bifurcation, black crosses in \textbf{a}), damped oscillations
below the gray curve \eqref{eq:principal_splitting} (gray cross in
\textbf{a}). Left: Oscillation frequency \eqref{eq:omega_d_crit}
at the Hopf bifurcation.\textbf{ (c)} Averaged cross covariance between
excitatory neurons and theory \eqref{eq:covariance_time} (black).
Simulated data averaged over $10^{6}$ neuron pairs for $100\sec$.
\textbf{(d) }Autocovariance of excitatory neurons ($\delta$-peak
not shown) averaged over $2500$ neurons for $100\sec$.\foreignlanguage{british}{
}\foreignlanguage{english}{}\label{fig:phase_diagram}}
\end{figure}

Global properties of the network dynamics can be inferred by considering
the spectrum of equation \eqref{eq:coherence_causal_A}, those complex
frequencies $z_{k}$ at which the expression has a pole due to the
function $U(z)$. These poles are resonant modes of the network, where
the real part $\Re(z_{k})$ denotes the damping of the mode, and the
imaginary part $\Im(z_{k})$ is the oscillation frequency. A pole
appears whenever $z_{k}$ is a single root of $U^{-1}(z_{k})=H^{-1}(-iz_{k})-L=0$.
With the Fourier representation $H(\omega)=\frac{e^{-i\omega d}}{1+i\omega\taue}$
of the response kernel \eqref{eq:imp_response}, the poles correspond
to the spectrum of the delay differential equation $\taue\frac{dy}{dt}(t)=-y(t)+Ly(t-d)$,
(cf. \cite{Guillouzic99_3970}) which describes the evolution of the
population averaged activity. As shown in \prettyref{app:spectrum},
the location of the poles can be expressed by the branches $k$ of
the Lambert $W$ function, the solution of $W_{k}e^{W_{k}}=x$ \cite{Corless96_329},
as
\begin{eqnarray}
z_{k} & = & -\frac{1}{\taue}+\frac{1}{d}W_{k}(L\frac{d}{\taue}e^{\frac{d}{\taue}})\quad k\in\mathbb{N}_{0}.\label{eq:pole_r}
\end{eqnarray}
The spectrum only depends on the population feedback $L$, the delay
$d$ and the effective time constant $\taue$ of the neural response
kernel. This explains why keeping $L$ constant while scaling the
network in \prettyref{fig:scaling_corr_functions} yields shape invariant
covariance functions. A typical spectrum is shown in \prettyref{fig:phase_diagram}a
as an inset, where each dot marks one of the poles $z_{k}$ \eqref{eq:pole_r}.
The two principal branches of $W_{k}$ are the modes with the largest
real part $\Re(z_{k})$, and hence with the least damping, dominating
the network dynamics. The remaining branches appear as conjugate pairs
and their real parts are more negative, corresponding to stronger
damping. Investigating the location of the principal branches therefore
enables us to classify the dynamics in the network. Their dependence
on the delay is shown in \prettyref{fig:phase_diagram}a as a parametric
plot in $d$. The point at which the two real principal solutions
turn into a complex conjugate pair marks a transition from purely
exponentially decaying dynamics to damped oscillations. This happens
at sufficiently strong negative coupling $L$ or sufficiently long
delay $d$, precisely when the argument of $W_{k}$ is smaller than
$-e^{-1}$ \cite{Corless96_329}, leading to the condition 
\begin{eqnarray}
L & < & -\frac{\taue}{d}e^{-\frac{d}{\taue}-1}.\label{eq:principal_splitting}
\end{eqnarray}
The gray cross marks this point in \prettyref{fig:phase_diagram}a,
the gray curve shows the corresponding relation of feedback and delay
in the phase diagram \prettyref{fig:phase_diagram}b. In the region
below the curve, the dominant mode of fluctuations in the network
is thus damped oscillatory, whereas above the curve fluctuations are
relaxing exponentially in time.

For sufficiently long delay $d$ the principal poles may assume positive
real values, leading to ongoing oscillations, a Hopf bifurcation.
The condition under which this happens can be derived from $H^{-1}(\omega_{\mathrm{crit.}})=L$,
as detailed in \prettyref{app:spectrum}. Equating the absolute values
on both sides leads to the condition $\omega_{\mathrm{crit.}}\taue=\sqrt{L^{2}-1}$:
oscillations can only be sustained, if the negative population feedback
is sufficiently strong $L<-1$. The oscillation frequency increases
the stronger the negative feedback. The condition for the phases leads
to the critical delay required for the onset of oscillations (see
\prettyref{app:spectrum} for details)
\begin{eqnarray}
\frac{d_{\mathrm{crit.}}}{\taue} & = & \frac{\pi-\arctan(\sqrt{L^{2}-1})}{\sqrt{L^{2}-1}}.\label{eq:d_crit_main}
\end{eqnarray}
This relation is shown as the black curve in the phase diagram \prettyref{fig:phase_diagram}b.
The oscillatory frequency on the bifurcation line, at the onset of
oscillations can be expressed as
\begin{eqnarray}
2\pi f_{\mathrm{crit.}}d=\omega_{\mathrm{crit.}}d & = & \pi-\arctan(\sqrt{L^{2}-1}),\label{eq:omega_d_crit}
\end{eqnarray}
which is shown in the left sub-panel of the phase diagram \prettyref{fig:phase_diagram}b.
Consequently, the oscillation frequency $f_{\mathrm{crit.}}$ at the
onset is between $(4d_{\mathrm{crit.}})^{-1}$ and $(2d_{\mathrm{crit.}})^{-1}$
, depending on the strength of the feedback $L$, approaching $f_{\mathrm{crit.}}=(4d_{\mathrm{crit.}})^{-1}$
at the onset with increasing negative feedback.

Changing the synaptic delay homogeneously for all synapses in the
network allows us to observe the transition of the network from exponentially
damped, to oscillatory damped, and finally to oscillatory dynamics.
For a short delay of $d=0.5\ms$ the dynamics is dominated by the
single real pole near $-1/\taue$ (brown dot in \prettyref{fig:phase_diagram}a)
and the covariance function is exponentially decaying (\prettyref{fig:phase_diagram}c).
Increasing the delay to $d=1\ms$ the principal poles split into a
complex conjugate pair as the delay crosses the gray curve in \prettyref{fig:phase_diagram}b
so that side troughs become visible in the covariance function in
\prettyref{fig:phase_diagram}c. Further increasing the delay, the
network approaches the point of oscillatory instability, where a Hopf
bifurcation occurs, marked by black crosses in \prettyref{fig:phase_diagram}a
and the black curve in \prettyref{fig:phase_diagram}b. The damping
of oscillations decreases as the system approaches the bifurcation
(\prettyref{fig:phase_diagram}c).  The structure of the auto covariance
function of single spike trains (\prettyref{fig:phase_diagram}d)
is dominated by the dip due to refractoriness of the neuron after
reset. At pronounced network oscillations, the autocorrelation function
shows a corresponding modulation. The neglect of the oscillating continuous
part of the autocorrelation function in the theory does apparently
not have a pronounced effect on the obtained cross correlation functions,
evidenced by the good agreement between direct simulation and theory
for $d=5\ms$. For weaker oscillations, e.g. at $d=3\ms$, the coherence
time of the oscillations is typically shorter than the width of the
dip in the autocorrelation. The phase coherence of the oscillation
is then shorter than the typical inter-spike-interval and single neuron
spike trains are irregular. This state is known as synchronous irregular
activity \cite{Brunel00_183}, where the activity of a single neuron
is irregular, but collectively the neurons participate in a global
oscillation.

\begin{figure}
\selectlanguage{british}%
\raggedleft{}\includegraphics[scale=0.8]{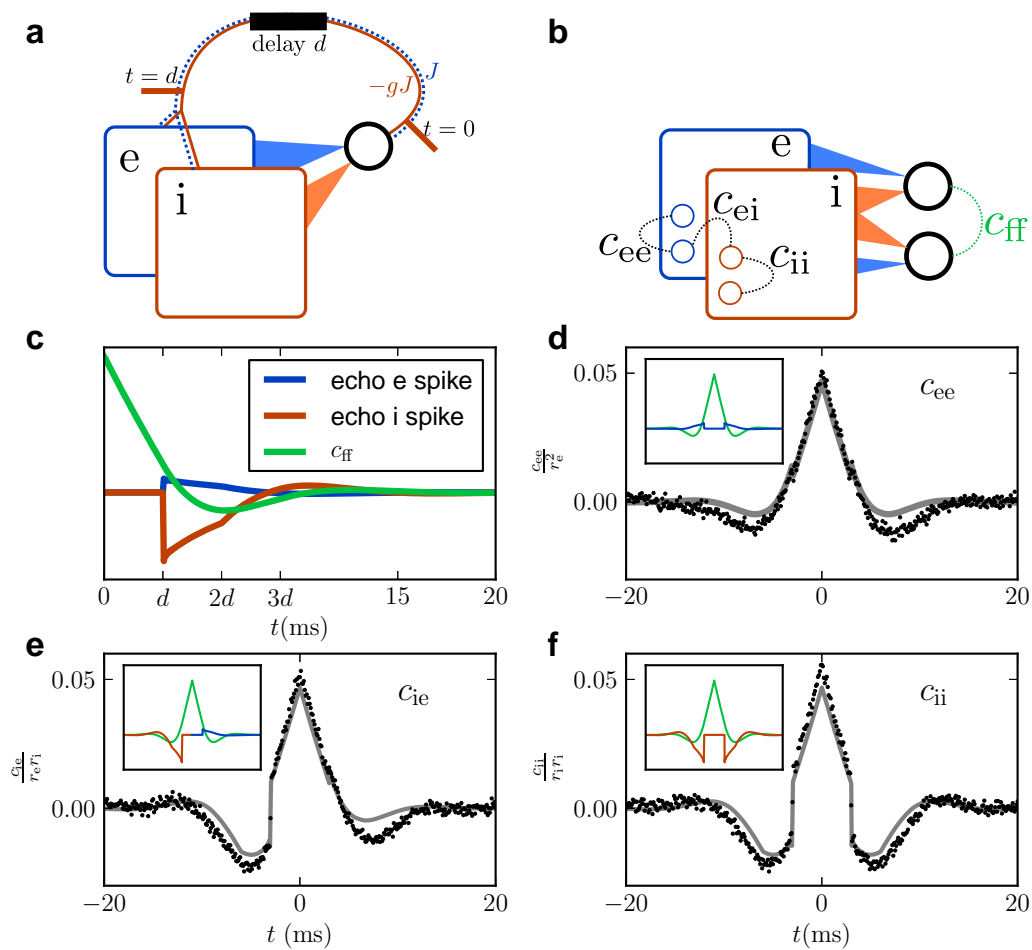}\foreignlanguage{american}{\caption{Composition of covariance functions. \textbf{(a)} Network echo caused
by a spike sent at time $t=0$ sets in after one synaptic delay ($d=3\ms$):
red curve in \textbf{c}, inhibitory spike, blue curve in \textbf{c},
excitatory spike. \textbf{(b)} Correlated inputs from the network
to a pair of neurons (black circles) cause covariance $c_{\mathrm{ff}}$
between their outputs (green curve in \textbf{c}). \textbf{(d)}-\textbf{(f)}
Covariance functions averaged over pairs of neurons (black dots,
\textbf{d}: excitatory pairs, \textbf{e}: excitatory-inhibitory pairs,
\textbf{f}: inhibitory pairs) and theory \eqref{eq:covariance_time}
(gray underlying curves). Inset shows the two components from \textbf{c}
that are added up.\foreignlanguage{british}{}\label{fig:corr_components}}
}\selectlanguage{american}%
\end{figure}

If the network is not in the oscillatory state, all modes are damped
in time, i.e. all poles $z_{k}$ of the function $U(z)$ appearing
in equation \eqref{eq:coherence_causal_A} lie in the left complex
half plane, $\Re(z_{k})<0$. Hence, the dynamics is stable, and we
can expect to obtain a unique solution for the covariance as an observable
of this stable dynamics. We perform the Fourier transform to time
domain using the residue theorem $u(t)=\frac{1}{2\pi i}\oint_{E_{t}}U(z)\, e^{zt}\, dz=\sum_{z_{k}\in E_{t}}\mathrm{Res}(U,z_{k})\, e^{z_{k}t}$,
where the integration path $E_{t}$ proceeds along the imaginary axis
from $-i\infty$ to $i\infty$ and is then closed in infinity in the
left half plane (for $t>d$) or the right half plane (for $0<t<d$)
to ensure convergence, resulting in (see \prettyref{app:back_trafo_residue}
for the detailed calculation)

\begin{eqnarray}
u(t) & = & \sum_{k=0}^{\infty}\frac{1}{\left(1+z_{k}\taue\right)d+\taue}\Theta(t-d)e^{z_{k}(t-d)}.\label{eq:u}
\end{eqnarray}
The back transform of $V(\omega)\defeq|U(\omega)|^{2}$ proceeds along
similar lines and results in

\begin{eqnarray}
v(t) & = & \sum_{k=0}^{\infty}\frac{1}{\left(1+z_{k}\taue\right)d+\taue}\,\frac{e^{z_{k}|t|}}{(1-z_{k}\taue)-Le^{z_{k}d}}.\label{eq:v}
\end{eqnarray}
The population averaged covariance functions in the time domain then
follow from equation \eqref{eq:coherence_causal_A} for $t>0$ as

\begin{eqnarray}
\c(t) & = & r\,\frac{Kw}{N}\left(\begin{array}{cc}
1 & -g\\
1 & -g
\end{array}\right)\; u(t)\nonumber \\
 & + & r\,\frac{(Kw)^{2}}{N}(1+g^{2}\gamma)\left(\begin{array}{cc}
1 & 1\\
1 & 1
\end{array}\right)\; v(t),\label{eq:covariance_time}
\end{eqnarray}
which is the central result of our work. \prettyref{fig:corr_components}
shows the comparison of the theory \eqref{eq:covariance_time} with
the covariance functions obtained by direct simulation. The analytical
expression unveils that the covariance functions are composed of two
components: the first line in equation \eqref{eq:covariance_time}
has heterogeneous matrix elements and hence depends on the neuron
types under consideration. Its origin is illustrated in \prettyref{fig:corr_components}a:
If one of the neurons emits a spike, as indicated, this impulse travels
along its axon and reaches the target neurons after one synaptic delay
$d$. Depending on the type of the source neuron, the impulse excites
(synaptic amplitude $J$) or inhibits ($-gJ$) its targets. Its effect
is therefore visible in the pairwise covariance function as a positive
(blue) or negative (red) deflection, respectively in \prettyref{fig:corr_components}c.
This deflection not only contains the direct synaptic effect, but
also infinitely many reverberations of the network, seen formally
in equation \eqref{eq:u}. This expression is not proportional to
the kernel $h(t)$ directly, but is rather a series including the
whole spectrum of the network. The shape of the spike echo consequently
shows onsets of reverberations at integer multiples of the synaptic
delay (\prettyref{fig:corr_components}c), being transmitted over
multiple synaptic connections. The contribution of the second line
in equation \eqref{eq:covariance_time} follows the intuitive argument
illustrated in \prettyref{fig:corr_components}b. The incoming activity
from the network to a pair of neurons is correlated. As the input
statistics is the same for each neuron, this contribution is identical
for any pair of neurons (green curve in \prettyref{fig:corr_components}c).
The sum of both components results in the covariance functions shown
in \prettyref{fig:corr_components}d-f. The same analytical solution
is shown for different delays in \prettyref{fig:phase_diagram}c showing
good agreement with direct simulation. For different sizes of simulated
networks in \prettyref{fig:scaling_corr_functions}b-d the analytical
expression \eqref{eq:covariance_time} explains why the spike echo
does not become negligible in the thermodynamic limit $N\rightarrow\infty$:
for fixed population feedback $L$, both contributions in equation
\eqref{eq:covariance_time} scale as $1/N$, so the relative contribution
of the echo stays the same. This also explains the apparent paradox
(see \prettyref{fig:correlation_transmission}), that covariance functions
in recurrent networks not only depend on the input statistics, but
in addition the spike feedback causes a reverberating echo. The power
spectrum of population-averages is dominated by pairwise covariances,
explaining the different spectra observed in the excitatory and inhibitory
population activity \cite{Kriener08_2185}. Scaling the network such
as to keep the marginal statistics of single neurons constant, $J\propto w\propto1/\sqrt{N}$
\cite{Vreeswijk96,Renart10_587} changes the spectrum \eqref{eq:pole_r},
because the feedback increases as $L\propto\sqrt{N}$ which can ultimately
lead to oscillations as shown in \prettyref{fig:phase_diagram}c.\foreignlanguage{british}{}
\begin{figure}
\selectlanguage{british}%
\raggedleft{}\includegraphics[scale=0.8]{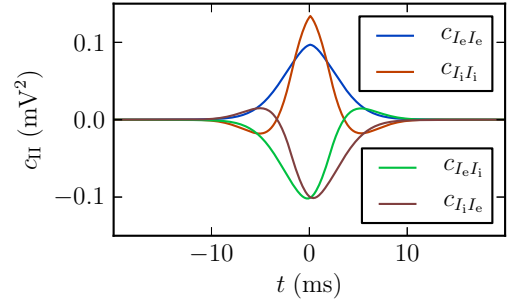}\caption{\selectlanguage{american}%
Covariance between synaptic currents of a pair of neurons in a recurrent
random network in analogy to in vivo experiments \cite{Okun_535_08}.
Covariance of excitatory contributions (blue, $c_{I_{\mathrm{e}}I_{\mathrm{e}}}=q\ast(J^{2}pKa_{\mathrm{e}}+J^{2}K^{2}c_{\mathrm{ee}})$),
analogously between inhibitory contributions (red), between excitatory
and inhibitory contribution (green, $c_{I_{\mathrm{e}}I_{\mathrm{i}}}=-q\ast(gJ{}^{2}\gamma K{}^{2}c_{\mathrm{ei}})$),
and $C_{I_{i}I_{e}}$ analogously (brown). Currents filter spiking
input by an exponential kernel with time constant $\taus=2\ms$ \eqref{eq:diffeq_iaf},
leading to the filtering of the covariances by $q(t)=\frac{\taum^{2}}{2\taus}e^{-|t|/\taus}$.
\foreignlanguage{british}{\label{fig:Okun_Lampl}}\selectlanguage{british}%
}
\selectlanguage{american}%
\end{figure}

\section{Discussion}

The present work qualitatively explains certain features of the correlation
structure of simultaneously recorded synaptic currents of two cells
in vivo. Novel experimental techniques are able to separate contributions
of excitatory and inhibitory inputs \cite{Okun_535_08}. We calculate
such covariances in a random network and show that the covariance
between synaptic impulses decomposes into a linear combination of
the covariance of the spiking activity and the autocovariance functions
(see caption of \prettyref{fig:Okun_Lampl}). Each synaptic impulse
has a certain time course, here modeled as a first-order low-pass
filter with time constant $\taus=2\ms$ (see \eqref{eq:diffeq_iaf}).
The covariances between these filtered currents are shown in \prettyref{fig:Okun_Lampl}.
Their temporal structure resembles those measured in cortex in vivo
\cite[their Figure 1e,f]{Okun_535_08}: covariances between afferents
of the same type are monophasic and positive, while the covariances
between excitatory and inhibitory afferents are biphasic and mostly
negative. The lag reported between inhibitory and excitatory activity
\cite[their Figure 2b]{Okun_535_08}, which was also observed in binary
random networks \cite{Ginzburg94,Renart10_587}, is explained by the
echo of the spike contributing to the covariance function. In contrast
to previous work, we take the delayed and pulsed synaptic interaction
into account. Without delays and with binary neurons \cite{Ginzburg94,Renart10_587}
the echo appears as a time lag of inhibition with respect to excitation.

Measurements of membrane potential fluctuations during quiet wakefulness
in the barrel cortex of mice \cite{Gentet10_422} showed that correlations
between inhibitory neurons are typically narrower than those between
two excitatory neurons \cite[their Figure 4A, 5B and Figure 5C,E]{Gentet10_422}.
These results qualitatively agree with our theory for covariances
between the spiking activity, because fluctuations of the membrane
potential are uniformly transferred to fluctuations of the instantaneous
firing intensity. The direct measures of spiking activity \cite[their Figure 6]{Gentet10_422}
confirm the asymmetric correlation between excitation and inhibition.
The low correlation between excitatory neurons reported in that study
may partly be due to the unnormalized, firing rate dependent measure
and the low rates of excitatory neurons.

The qualitative features of the cross correlation functions, namely
their different widths and their asymmetry for excitatory-inhibitory
pairs, are generic and agree with experimental results. They are fully
explained by the decomposition into an echo term and a term corresponding
to the feed-forward transmission of correlation. This decomposition
merely relies on the fact that the autocorrelation of a spike train
has a delta peak and that a spike triggers a response in the target
cell with positive or negative sign for excitatory and inhibitory
connections, respectively. Hence, we expect that these features are
robust and survive also for more realistic network models with many
heterogeneous subpopulations. For weakly correlated fluctuations,
if the input to each cell is sufficiently noisy, a linear approximation
of the neuronal response provides a viable first order approximation
also for non-linear neuron dynamics, as shown here. We suspect that
a deeper reason why such a linearization is possible is the inherent
decorrelation \cite{Tetzlaff12_e1002596} by negative feedback in
networks in the inhibition dominated regime. The decorrelation keeps
population fluctuations small and hence prevents strong excursions
that would exceed the validity of the linear approximation.

Oscillations in the $\gamma$ range ($25-100\Hz$) are ubiquitous
in population measures of neural activity in humans, and have earlier
been explained in networks of leaky integrate-and-fire model neurons
\cite{Brunel00_183} by the Hopf bifurcation induced by delayed negative
feedback. For the regime of high noise we here uncover a simpler analytical
condition for the onset \eqref{eq:d_crit_main} and frequency \eqref{eq:omega_d_crit}
of fast global oscillations. For lower noise, deviations of the non-linear
leaky integrate-and-fire dynamics from the linear theory presented
here are expected.

A traditional motivation to scale the network size to infinity is
to obtain an analytic solution for an otherwise hard problem. Biologically
realistic networks have a finite size. In the present work we have
determined the correlation structure for such finite sized networks
analytically. We present the scaling of the correlation functions
in \prettyref{fig:scaling_corr_functions} to relate our work to previous
results that applied scaling arguments to obtain the correlation structure
\cite{Renart10_587}. The latter work investigated networks of binary
neurons without conduction delays and assumed a scaling of the synaptic
amplitudes $J\propto1/\sqrt{N}$. Such a scaling increases the overall
feedback strength $L\propto pNw\propto\sqrt{N}$. This has two consequences:
Firstly, as seen from \eqref{eq:corrtrans_theory}, the relative contributions
of the spike echo and the feed forward term change with $L$ and hence
with network size. Therefore the shape of correlation functions depends
on the network size. Secondly, the overall dynamic regime of the network
is affected by the scaling. This can be seen from \prettyref{fig:phase_diagram}b.
For non-zero synaptic conduction delays, the network eventually becomes
oscillatory if the network size exceeds a certain value. This happens
precisely at the point where $L$ crosses the bifurcation line shown
in \prettyref{fig:phase_diagram}b. We therefore propose an alternative
scaling $J\propto1/N$ here for which we show that it preserves the
shape of correlation functions and the overall network state. Only
the magnitude of the cross correlation functions decreases $\propto1/N$.
However, a caveat of this scaling is that while it preserves the global
network properties, it affects the working point of each individual
neuron, because the fluctuations due to the local synaptic input decrease
$\propto1/\sqrt{N}$. We alleviated this shortcoming by supplying
each neuron with additional, uncorrelated noise.

Our results are based on a simplified network model composed of two
homogeneous (excitatory and inhibitory) subpopulations of leaky integrate-and-fire
neurons. The real cortex, in contrast, is highly heterogeneous in
several respects: Its layered structure with layer-specific neuron
and connection properties (e.g. time constants, spike thresholds,
synaptic weights, in-degrees) requires the distinction of more than
two subpopulations. Even within each subpopulation and for each combination
of subpopulations, the neuron and connection parameters are not constant
but broadly distributed. Further, a plethora of cortical neurons,
in particular various types of interneurons, exhibit a much richer
dynamical repertoire (e.g. resonating behavior) than the leaky integrate-and-fire
neuron model (regular spiking integrator). In principle, the mathematical
framework presented in this article can be extended to networks composed
of $n$ ($n>2$) heterogeneous subpopulations of different neuron
types. This would require to account for subpopulation-specific working
points (e.g. due to layer-specific firing rates; see \cite{Potjans12_358})
and for the effect of parameter distributions \cite{Tetzlaff09_258,Roxin11_8}
and the single-neuron dynamics on the effective linearized subpopulation
responses. This extended theory would result in an $n$-dimensional
linear algebraic equation for the subpopulation-averaged cross spectra,
similar to \eqref{eq:coherence_causal_A} where $n=2$. For $n>2$,
this equation most likely needs to be solved numerically.

A fundamental assumption of the presented theory is that the network
states show irregular single neuron dynamics. This requirement arises
from the analytical description replacing spike trains by spike densities
and a stochastic realization of spikes. Regular spike trains are outside
the scope of such a description. Moreover, the approximation of the
neuronal response to linear order is only a viable approach in sufficiently
asynchronous network states with low correlations. States with stronger
correlations, such as observed in convergent-divergent feed-forward
structures \cite{Abeles91,Diesmann99_529}, require an explicit treatment
of the non-linear response \cite{Goedeke08_015007}.

From a physics viewpoint, neuronal networks unite several interesting
properties. They do not reach thermodynamic equilibrium even in the
stationary state, as detailed balance does not hold for all pairs
of states of the system. In detailed balance, the rate of transition
from one state to another is equal to the rate of the reverse transition.
For a leaky integrate-and-fire neuron the state of the neuron is uniquely
determined by its membrane voltage. In the stationary state neurons
fire with a constant rate, so there is a continuous flux of the neurons'
voltage from reset up to threshold. Imagining a discretization of
the voltage axis we see that a pair of adjacent voltage-intervals
between reset and threshold is more often traversed from lower to
higher voltage than in the reverse direction, so obviously detailed
balance does not hold even for a single neuron in the stationary state.
Moreover, the interaction between pairs of neurons is directed, delayed,
pulsed, and depends on the flux of the sending neuron's state variable
at threshold. In contrast, pairwise interactions frequently studied
in physics, like the Coulomb interaction or exchange interaction,
can be expressed by a pair potential and are thus symmetric (undirected),
instantaneous, and depend directly on the state variables (e.g. spatial
coordinates or spins) of the pair of interacting particles. Non-equilibrium
systems are at the heart of ubiquitous transport phenomena, like heat
or electric conduction. Understanding fluctuations in such a system
marks the starting point to infer macroscopic properties by the assertion
of the fluctuation-dissipation theorem that connects microscopic fluctuations
to macroscopic transport properties. Despite the non-equilibrium dynamics
and the non-conservative pairwise interaction, in this manuscript
we develop a simple analytical framework merely based on linear perturbation
theory that explains time dependent covariance functions of the activity
of pairs of integrate-and-fire model neurons in a recurrent random
network. Formally our approach resembles the step from the kinetic
Ising model near equilibrium to its non-equilibrium counterpart, the
network of binary neurons \cite{Ginzburg94}. A difference is the
spiking interaction considered in our work, which led us to the describe
each neuron in terms of the flux over threshold (spike train) rather
than by its state variables (membrane voltage and synaptic current).
In this respect, we follow the established mean-field approach for
spiking neuronal systems \cite{Amit97,Brunel00_183}. However, while
this mean field approach proceeds by assuming vanishing correlations
to obtain the dynamics of the ensemble averaged activity, we here
derive and solve the self-consistency equation for the pairwise averaged
covariances of the microscopic system.

The typical time scale of covariance functions found here coincides
with the time window of biological synaptic plasticity rules \cite{Morrison08_459},
so that non-trivial interactions of dynamics and structure are expected.
It is our hope that the novel capability to resolve the temporal structure
of covariances in spiking networks presented here proves useful as
a formal framework to further advance the theory of these correlated
non-equilibrium systems and in particular serves as a further stepping
stone in the endeavor to understand how learning on the system level
is implemented by the interplay of neuronal dynamics and synaptic
plasticity.

\ack{}{}

We are thankful to Birgit Kriener, Sonja Gr{\"u}n and George Gerstein
for discussions and exploratory simulations in fall 2005 that led
to the observation of asymmetric covariance functions. Partially supported
by the Helmholtz Association: HASB and portfolio theme SMHB, the Next-Generation
Supercomputer Project of MEXT, EU Grant 15879 (FACETS), EU Grant 269921
(BrainScaleS). All network simulations carried out with NEST (http://www.nest-initiative.org).

\appendix

\section{Response kernel of the LIF model\label{app:response_kernel}}

The response kernel kernel $h_{ij}$ needs to be related to the dynamics
of the neuron model \prettyref{eq:diffeq_iaf}. Here we present an
approximation of this kernel which is sufficiently accurate to allow
quantitative predictions, but yet simple enough to enable an analytical
solution for the correlation structure. If the synaptic time constant
is short $\taus\ll\taum$, the synaptic amplitude $J$ can be thought
of as the amplitude of the jump in the membrane potential $V$ caused
upon arrival of an incoming impulse. If correlations between incoming
spike trains are sufficiently small, the first and second moments
of the summed impulses $\taum\sum_{j}J_{ij}s_{j}(t-d)$ are $\mu_{i}=\taum\sum_{j}J_{ij}r_{j}$
and $\sigma_{i}^{2}=\taum\sum_{j}J_{ij}^{2}r_{j}$, respectively,
if the inputs' statistics can be approximated by Poisson processes
of rate $r_{j}$ each. For small $J$ and a high total rate, the system
of differential equations \prettyref{eq:diffeq_iaf} is hence approximated
by a stochastic differential equation driven by a unit variance Gaussian
white noise $\xi$

\begin{eqnarray*}
\taum\frac{dV_{i}}{dt} & = & -V_{i}+I_{s,i}(t)\\
\taus\frac{dI_{i}}{dt} & = & -I_{i}+\mu_{i}+\sigma_{i}\sqrt{\taum}\xi(t).
\end{eqnarray*}
The stationary firing rate in this limit is given by \cite{Fourcaud02}
\begin{eqnarray}
r_{i}^{-1} & = & \taur+\taum\sqrt{\pi}\left(F(y_{\theta})-F(y_{r})\right)\label{eq:rate}\\
f(y) & = & e^{y^{2}}(1+\mathrm{erf}(y))\quad F(y)=\int^{y}f(y)\, dy\nonumber \\
\text{with }y_{\theta,r} & = & \frac{V_{\theta,r}-\mu_{i}}{\sigma_{i}}+\frac{\alpha}{2}\sqrt{\frac{\taus}{\taum}}\quad\alpha=\sqrt{2}|\zeta(\frac{1}{2})|,\nonumber 
\end{eqnarray}
with Riemann's zeta function $\zeta$. The rate $r_{i}$ is the density
of action potentials per time. The response of the firing density
of the neuron $i$ at time point $t$ with respect to a point-like
deflection of the afferent input $s_{j}$ at time point $t^{\prime}$
defines the response kernel as the functional derivative 
\begin{eqnarray*}
\left\langle \frac{\delta s_{i}(t)}{\delta s_{j}(t^{\prime})}\right\rangle _{\s\backslash s_{j}} & = & \lim_{\epsilon\rightarrow0}\frac{1}{\epsilon}\left\langle s_{i}(t,\{\s(\tau)+\epsilon\delta(\tau-t^{\prime})\e_{j}|\tau<t\})\right.\\
 &  & -\left.s_{i}(t,\{\s(\tau)|\tau<t\})\right\rangle _{\s\backslash s_{j}}\\
 & \defeq & h_{ij}(t-t^{\prime})=w_{ij}\; h(t-t^{\prime}).
\end{eqnarray*}
Here we used the homogeneity, namely the identical input statistics
of each neuron $i$, leading to the same temporal shape $h(t)$ independent
of $i$ and the stationarity, so that the kernel only depends on the
time difference $t-t^{\prime}$. We choose $h(t)$ to have unit integral
and define $w_{ij}$ as the integral of the kernel. We determine the
temporal integral of the kernel as
\begin{eqnarray}
w_{ij} & = & \int_{-\infty}^{\infty}h_{ij}(t)\; dt\label{eq:dc_susz}\\
 & = & \frac{\partial r_{i}}{\partial r_{j}}.\nonumber 
\end{eqnarray}
The second equality holds because the integral of the impulse response
equals the step response \cite{Oppenheim96}. Further, a step in the
density $s_{j}$ corresponds to a step of $r_{j}$. Up to linear approximation
the effect of the step in the rate $r_{j}$ on the rate $r_{i}$ can
be expressed by the derivative considering the perturbation of the
mean $\mu_{i}$ and the variance $\sigma_{i}^{2}$ upon change of
$r_{j}$. Using equation \prettyref{eq:rate} we note that by chain
rule $\frac{\partial r_{i}}{\partial r_{j}}=-r_{i}^{2}\frac{\partial r_{i}^{-1}}{\partial r_{j}}$.
The latter derivative follows as $\frac{\partial r_{i}^{-1}}{\partial r_{j}}=\frac{\partial r_{i}^{-1}}{\partial y_{\theta}}\frac{\partial y_{\theta}}{\partial r_{j}}+\frac{\partial r_{i}^{-1}}{\partial y_{r}}\frac{\partial y_{r}}{\partial r_{j}}$.
The first derivative in both terms yields $\frac{\partial r_{i}^{-1}}{\partial y_{A}}=\sqrt{\pi}\taum f(y_{A})$
with $y_{A}\in\{y_{\theta},y_{r}\}$. The second derivative evaluates
with $y_{A}=\frac{A-\mu_{i}}{\sigma_{i}}+\frac{\alpha}{2}\sqrt{\frac{\taus}{\taum}}$
to $\frac{\partial y_{A}}{\partial r_{i}}=-\frac{1}{\sigma_{i}}\taum J_{ij}-\frac{A-\mu_{i}}{\sigma_{i}^{2}}\frac{\taum J_{ij}^{2}}{2\sigma_{i}}=-\taum\frac{J_{ij}}{\sigma_{i}}(1+\frac{A-\mu_{i}}{\sigma_{i}}\frac{J_{ij}}{2\sigma_{i}})$.
So together we obtain
\begin{eqnarray}
w_{ij} & = & \alpha J_{ij}+\beta J_{ij}^{2}\label{eq:w_ij}\\
\text{with }\alpha & = & \sqrt{\pi}(\taum r_{i})^{2}\frac{1}{\sigma_{i}}\left(f(y_{\theta})-f(y_{r})\right)\nonumber \\
\text{and }\text{\ensuremath{\beta}} & = & \sqrt{\pi}(\taum r_{i})^{2}\left(f(y_{\theta})\,\frac{V_{\theta}-\mu_{i}}{2\sigma_{i}^{3}}-f(y_{r})\,\frac{V_{r}-\mu_{i}}{2\sigma_{i}^{3}}\right).\nonumber 
\end{eqnarray}

\section{Cross spectral matrix in frequency domain\label{app:covariance_general_autocorrelation}}

The autocovariance $A(\omega)$ in the frequency domain ($F(\omega)=\Fourier[f](\omega)=\int_{-\infty}^{\infty}f(t)e^{-i\omega t}\, dt$)
has two different terms. The first term is a constant $r$ due to
the spiking with rate $r$ resulting from the delta peak $r\delta(t)$
in time domain. The second term is the continuous function $a_{\text{c}}(t)$,
for example due to refractoriness of the neuron. Further follows from
$a(t)=a(-t)$ that $A(\omega)=A(-\omega)$. For $\tau>0$ the covariance
matrix fulfills the linear convolution equation \eqref{eq:integral_equation_corrfunction_avg}.
As this equation only holds for the positive half of the time axis,
we cannot just apply the Fourier transform to obtain the solution.
For negative time lags $\tau<0$ the covariance matrix is determined
by the symmetry $\c(\tau)=\c^{T}(-\tau)$. Here we closely follow
\cite{Hawkes71_438} and employ Wiener-Hopf theory \cite{Hazewinkel02}
to derive an equation for the cross spectral matrix in the frequency
domain that has the desired symmetry and solves \prettyref{eq:integral_equation_corrfunction_avg}
simultaneously. To this end we introduce the auxiliary matrix $\b(\tau)=\left(h\ast(\M\c)\right)(\tau)+\Q(h\ast a)(\tau)-\c(\tau)$
for $-\infty<\tau<\infty$. Obviously, $\b(\tau)=0$ for $\tau>0$.
Since the defining equation for $\b$ holds on the whole time axis,
we may apply the Fourier transform to obtain $\B(\omega)=H(\omega)\left(\M\C(\omega)+\Q A(\omega)\right)-\C(\omega)$.
Solving for $\C$ 
\begin{eqnarray}
\C(\omega) & = & (\I-\M H(\omega))^{-1}\left(H(\omega)A(\omega)\Q-\B(\omega)\right)\label{eq:C_solved}
\end{eqnarray}
and using the symmetry $\C(\omega)=\C^{T}(-\omega)$ we obtain the
equation
\begin{eqnarray*}
 &  & \left(H(\omega)A(\omega)\Q-\B(\omega)\right)\left(\I-\M^{T}H(-\omega)\right)\\
 & = & \left(\I-\M H(\omega)\right)\left(H(-\omega)A(-\omega)\Q^{T}-\B^{T}(-\omega)\right).
\end{eqnarray*}
We observe that $\Q\M^{T}=\M\Q^{T}$ is symmetric and with $A(\omega)=A(-\omega)$
the term proportional to $|H(\omega)|^{2}$ cancels on both sides,
remaining with
\begin{eqnarray}
 &  & H(\omega)A(\omega)\Q+\left(\I-\M H(\omega)\right)\B^{T}(-\omega)\label{eq:pre_B}\\
 & = & H(-\omega)A(\omega)\Q^{T}+\B(\omega)\left(\I-\M^{T}H(-\omega)\right).\nonumber 
\end{eqnarray}
We next introduce $D(\omega)=H(\omega)A(\omega)$ which we split into
$D(\omega)=D_{+}(\omega)+D_{-}(\omega)$, chosen such that $d_{+}(t)$
(in time domain) vanishes for $t<0$ and $d_{-}(t)$ vanishes for
$t>0$. Consequently the Fourier transforms of both terms may have
poles in distinct complex half-planes: $D_{+}(\omega)$ may only have
poles in the upper half plane $\Im(\omega)>0$ and the function vanishes
for $\lim_{|\omega|\rightarrow\infty,\Im(\omega)<0}D_{+}(\omega)=0$,
following from the definition of the Fourier integral. For $D_{-}(\omega)$
the half planes are reversed. The analytical properties of $H(\omega)$
are thus similar to those of $D_{+}(\omega)$, those of $B(\omega)$
are similar to $D_{-}(\omega)$. We sort the terms in \prettyref{eq:pre_B}
such that the left hand side only contains terms that vanish at infinity
in the lower half plane $\Im(\omega)<0$, the right hand side those
that vanish in infinity in the upper half plane $\Im(\omega)>0$
\begin{eqnarray}
 &  & D_{+}(\omega)\Q-D_{-}(-\omega)\Q^{T}+\left(\I-\M H(\omega)\right)\B^{T}(-\omega)\label{eq:analytic_continuation}\\
 & = & D_{+}(-\omega)\Q^{T}-D_{-}(\omega)\Q+\B(\omega)\left(\I-\M^{T}H(-\omega)\right).\nonumber 
\end{eqnarray}
The left hand side consequently is analytic for $\Im(\omega)<0$ the
right hand side is analytic for $\Im(\omega)>0$, so \eqref{eq:analytic_continuation}
defines a function that is analytic on the whole complex plane and
that vanishes at the border for $|\omega|\rightarrow\infty$. Hence
by Liouville's theorem it is $0$ and we can solve the right hand
side of \eqref{eq:analytic_continuation} for $\B$

\begin{eqnarray*}
\B(\omega) & = & \left(D_{-}(\omega)\Q-D_{+}(-\omega)\Q^{T}\right)\left(\I-\M^{T}H(-\omega)\right)^{-1}.
\end{eqnarray*}
Inserted into \prettyref{eq:C_solved} this yields with the definition
$\P(\omega)=\left(\I-\M H(\omega)\right)^{-1}$

\begin{eqnarray*}
\C(\omega) & = & \P(\omega)\left(H(\omega)A(\omega)\Q-\left(D_{-}(\omega)\Q-D_{+}(-\omega)\Q^{T}\right)\P^{T}(-\omega)\right)\\
 & = & \P(\omega)\left(\left(D_{+}(\omega)+D_{-}(\omega)\right)\Q(\I-\M^{T}H(-\omega))\right.\\
 &  & \phantom{\P(\omega)}\left.-D_{-}(\omega)\Q+D_{+}(-\omega)\Q^{T}\right)\P^{T}(-\omega)\\
 & = & \P(\omega)\left(D_{+}(\omega)\Q+D_{+}(-\omega)\Q^{T}-A(\omega)|H(\omega)|^{2}\Q\M^{T}\right)\P^{T}(-\omega).
\end{eqnarray*}
The latter expression can be brought to the form

\begin{eqnarray}
\C(\omega) & = & D_{+}(\omega)\P(\omega)\Q+D_{+}(-\omega)\Q^{T}\P^{T}(-\omega)\nonumber \\
 &  & +\left(D_{+}(\omega)H(-\omega)+D_{+}(-\omega)H(\omega)-A(\omega)|H(\omega)|^{2}\right)\times\label{eq:C_omega_sort_propagators}\\
 &  & \times\P(\omega)\M\Q^{T}\P^{T}(-\omega),\nonumber 
\end{eqnarray}
which has the advantage that the first two terms have poles in distinct
half planes $\Im(\omega)>0$, and $\Im(\omega)<0$, respectively.
This means these terms only contribute for positive and negative times,
respectively, the last term contributes for positive and negative
times.

\section{Spectrum of the propagator\label{app:spectrum}}

With the Fourier representation $H(\omega)=\frac{e^{-i\omega d}}{1+i\omega\taue}$
of the delayed exponential kernel \eqref{eq:imp_response} the averaged
cross spectrum \eqref{eq:coherence_causal_A} contains the two functions
$U(z)$ and $V(z)=|U(z)|^{2}$ defined on the complex frequency plane
$z=i\omega$. These functions may exhibit poles. The function $U$
has a pole $z_{k}$ whenever the denominator has a single root $H^{-1}(-iz)-L=0$
which amounts to the condition $(1+z_{k}\taue)e^{z_{k}d}=L$. These
complex frequencies can be expressed by the Lambert $W$ function,
the solution of $We^{W}=x$ \cite{Corless96_329}, by

\begin{eqnarray}
(1+z_{k}\taue)e^{z_{k}d} & = & L\label{eq:pole_condition}\\
(\frac{d}{\taue}+z_{k}d)e^{\frac{d}{\taue}+z_{k}d} & = & L\frac{d}{\taue}e^{\frac{d}{\taue}}\nonumber 
\end{eqnarray}
as

\begin{eqnarray*}
z_{k} & = & -\frac{1}{\taue}+\frac{1}{d}W_{k}(L\frac{d}{\taue}e^{\frac{d}{\taue}}),
\end{eqnarray*}
leading to \eqref{eq:pole_r}. The Lambert $W_{k}(x)$ function has
infinitely many branches $k$ \cite{Corless96_329}. The principal
branch has two real solutions, if $x>-e^{-1}$. The remaining branches
appear in conjugate pairs. For $x<-e^{-1}$ the principal solutions
turn into a complex conjugate pair. This happens at sufficiently strong
negative coupling $L$ or long delays $d$
\begin{eqnarray*}
L\frac{d}{\taue}e^{\frac{d}{\taue}} & < & -e^{-1}\\
L & < & -\frac{\taue}{d}e^{-\frac{d}{\taue}-1}\\
\text{or } & \frac{d}{\taue} & >W_{0}(-\frac{1}{Le}).
\end{eqnarray*}
The principal poles may assume positive real values, leading to oscillations.
The condition under which this happens can be derived from \eqref{eq:pole_condition}.
At the point of transition the pole can be written as $z=i\omega_{\mathrm{crit.}}$;
it is a solution to $(1+i\omega_{\mathrm{crit.}}\taue)e^{i\omega_{\mathrm{crit.}}d}=L.$
In order for this equation to be fulfilled, the absolute value and
the phase must be identical on both sides. The equation for the absolute
value requires $1+(\omega_{\mathrm{crit.}}\taue)^{2}=L^{2}$. This
means there are only oscillatory solutions, if the magnitude of the
feedback exceeds unity $L<-1$. Since the poles come in conjugate
pairs, we can assume w.l.o.g. that $\omega_{\mathrm{crit.}}>0$. The
condition for the absolute value hence reads

\begin{eqnarray}
\omega_{\mathrm{crit.}}\taue & = & \sqrt{L^{2}-1}.\label{eq:omega_crit}
\end{eqnarray}
This is the frequency of oscillation at the onset of the Hopf bifurcation.
For strong feedback $|L|\gg1$ the frequency increases linearly with
the magnitude of the feedback. The condition for the agreement of
the phase angles reads $\angle(1+i\omega_{\mathrm{crit.}}\taue)+\omega_{\mathrm{crit.}}d=0$,
so $\frac{\Im(1+i\omega_{\mathrm{crit.}}\taue)}{\Re(1+i\omega_{\mathrm{crit.}}\taue)}=-\frac{\Im(e^{i\omega_{\mathrm{crit.}}d})}{\Re(e^{i\omega_{\mathrm{crit.}}d})}=\tan\omega_{\mathrm{crit.}}d$,
which leads to $\tan\omega_{\mathrm{crit.}}d=-\omega_{\mathrm{crit.}}\taue.$
This equation has a solution in $\frac{\pi}{2}\le\omega_{\mathrm{crit.}}d\le\pi$.
In the limit of vanishing delay $d\rightarrow0$ the frequency goes
to infinity, as the solution converges to $\omega_{\mathrm{crit.}}d=\frac{\pi}{2}$.
This corresponds to the frequency $f_{\mathrm{crit.}}=\frac{1}{4d}$.
Inserting \prettyref{eq:omega_crit} leads to $\tan(\frac{d}{\taue}\sqrt{L^{2}-1})=-\sqrt{L^{2}-1}$,
which can be solved for the critical delay 
\begin{eqnarray}
\frac{d}{\taue} & = & \frac{\pi-\arctan(\sqrt{L^{2}-1})}{\sqrt{L^{2}-1}}\label{eq:d_crit}
\end{eqnarray}
where we took care that the argument of the tangent is in $[\frac{\pi}{2},\pi]$.
So with \eqref{eq:omega_crit} and \eqref{eq:d_crit} the oscillatory
frequency at the transition can be related to the synaptic delay as
\begin{eqnarray*}
2\pi f_{\mathrm{crit.}}d=\omega d & = & \pi-\arctan(\sqrt{L^{2}-1}).
\end{eqnarray*}

\section{Back transform by residue theorem\label{app:back_trafo_residue}}

In the non-oscillatory state all poles $z_{k}$ \prettyref{eq:pole_r}
have a negative real part. The function $U(z)=((1+z\taue)e^{zd}-L)^{-1}$
in \eqref{eq:coherence_causal_A} then has all poles in the left complex
half plane, $\Re(z_{k})<0\quad\forall k$. We perform the Fourier
back transform
\begin{eqnarray}
u(t) & = & \frac{1}{2\pi}\int_{-\infty}^{\infty}U(i\omega)\, e^{i\omega t}\, d\omega\label{eq:residue_backtrafo}\\
 & = & \frac{1}{2\pi i}\oint_{E_{t}}U(z)\, e^{zt}\, dz,\nonumber 
\end{eqnarray}
replacing the integration path by a closed contour $E_{t}$ following
the imaginary axis from $-i\infty$ to $i\infty$. In order to ensure
convergence of the integral, for $t<d$ we need $\Re(z)>0$, so we
close $E_{t<d}$ in infinity within the right half-plane, where the
integrand vanishes. Since there are no poles in the right half-plane,
for $t<d$ the path $E_{t<d}$ does not enclose any poles, so $u(t)=0$.
For $t\ge d$ the path $E_{t\ge d}$ must be closed in the left half-plane
to ensure convergence of \eqref{eq:residue_backtrafo}, so the residue
theorem yields
\begin{eqnarray}
u(t) & = & \Theta(t-d)\sum_{z_{k}\in E_{t>d}}\mathrm{Res}(U,z_{k})\, e^{z_{k}t}.\label{eq:residue_sum_backtrafo}
\end{eqnarray}
The residue can be calculated by linearizing the denominator of $U(z_{k}+z)$
around $z_{k}$

\begin{eqnarray*}
(1+(z_{k}+z)\taue)e^{(z_{k}+z)d}-L & = & (1+z_{k}\taue)e^{z_{k}d}e^{zd}-L+z\taue e^{z_{k}d}e^{zd}\\
 & = & L(e^{zd}-1)+z\taue e^{z_{k}d}(1+zd)+O(z^{2})\\
 & = & z(Ld+\taue e^{z_{k}d})+O(z^{2}),
\end{eqnarray*}
 which yields 
\begin{eqnarray*}
\mathrm{Res}(U,z_{k}) & = & \lim_{z\rightarrow0}zU(z_{k}+z)\\
 & = & \lim_{z\rightarrow0}\frac{z}{z(Ld+\taue e^{z_{k}d})}=\frac{1}{Ld+\taue e^{z_{k}d}}\\
 & = & \frac{e^{-z_{k}d}}{(1+z_{k}\taue)d+\taue},
\end{eqnarray*}
where we used \eqref{eq:pole_condition} in the last step. The poles
of $V(z)=U(z)U(-z)$ are located in both half-planes, consequently
$v(t)$ is nonzero on the whole time axis. Here we only calculate
$v(t)$ for positive times $t>0$, because it follows for negative
times by symmetry $v(-t)=v(t)$. The path has to be closed in the
left half-plane, where the poles $z_{k}$ have the residues
\begin{eqnarray*}
\mathrm{Res}(V,z_{k})=\mathrm{Res}(U,z_{k})U(-z_{k}) & = & \frac{1}{\left(1+z_{k}\taue\right)d+\taue}\;\frac{1}{(1-z_{k}\taue)-Le^{z_{k}d}}.
\end{eqnarray*}
So applying \eqref{eq:residue_sum_backtrafo} the functions $u$ and
$v$ are

\begin{eqnarray}
u(t) & = & \sum_{_{k=0}}^{\infty}\frac{1}{\left(1+z_{k}\taue\right)d+\taue}\Theta(t-d)e^{z_{k}(t-d)}\label{eq:u_v_sum}\\
v(t) & = & \sum_{k=0}^{\infty}\mathrm{Res}(V,z_{k})\left(\Theta(t)e^{z_{k}t}+\Theta(-t)e^{-z_{k}t}\right)\nonumber \\
 & = & \sum_{k=0}^{\infty}\frac{1}{\left(1+z_{k}\taue\right)d+\taue}\,\frac{1}{(1-z_{k}\taue)-Le^{z_{k}d}}e^{z_{k}|t|}.\nonumber 
\end{eqnarray}
The amplitude of the modes decrease with $k$. For all figures in
the manuscript we truncated the series after $k=30$.

\section{Simulation parameters used for figures\label{app:simulation_parameters}}

All network simulations were performed using NEST \cite{Gewaltig_07_11204}.
The parameters of the leaky integrate-and-fire neuron model \eqref{eq:diffeq_iaf}
throughout this work are $\taum=20\ms$,\foreignlanguage{english}{
$\taus=2\ms$, $\taur=2\ms$\textbf{, $V_{\theta}=15\mV$}, $V_{r}=0\mV$}.
All simulations are performed with precise spike timing and time stepping
of $0.1\ms$ \cite{Hanuschkin10_113}.\textbf{ }\prettyref{fig:correlation_transmission}
and \prettyref{fig:scaling_corr_functions}-\prettyref{fig:Okun_Lampl}
of the main text all consider recurrent random networks of $N$ excitatory
and $\gamma N$ inhibitory leaky integrate-and-fire model neurons
receiving input from randomly drawn neurons in the network and external
excitatory and inhibitory Poisson input, so that the first and second
moments are $\mu_{i}=15\mV$ and $\sigma_{i}^{2}=10\mV$, respectively.
Unless stated explicitly, we use $N(1+\gamma)=10000$ neurons except
in \prettyref{fig:scaling_corr_functions} where the number of neurons
is given in the legend. Each neuron has $K=pN$ incoming excitatory
synapses with synaptic amplitude $J$ independently and randomly drawn
from the pool of excitatory neurons, and $\gamma K=\gamma pN$ inhibitory
synapses with amplitude $-gJ$ (homogeneous Erd\H{o}s-R\'{e}nyi
random network with fixed in-degree), realizing a connection probability
$p=0.1$. Cross covariance functions are measured throughout as the
covariance between two disjoint populations of $1000$ neurons each
taken from the indicated populations in the network. Correlation functions
are evaluated with a time resolution of $0.1\ms$.

In \prettyref{fig:correlation_transmission} we use a synaptic delay
$d=1\ms$.

In \prettyref{fig:scaling_corr_functions} we keep the feedback of
the population rate constant $L=Kw(1-\gamma g)=\text{const.}$ Increasing
the size of the network $N$ the synaptic amplitude $J$ (which is
proportional to $w$ in linear approximation) needs to scale as $J=J_{0}N_{0}/N$,
where we chose $J_{0}=0.1\mV$ and $N_{0}=10000$ here. The variance
caused by local input from the network then decreases with increasing
network size $\propto1/N$, while the local mean is constant because
$L=\mathrm{const.}$ Each cell receives in addition uncorrelated external
balanced Poisson input, adjusted to keep the mean $\mu_{i}=15\mV$,
and fluctuations $\sigma_{i}=10\mV$ constant. This is achieved by
choosing the rates of the external excitatory ($r_{\mathrm{e},\ext}$,
amplitude $J_{\ext}=0.1\mV$) and inhibitory ($r_{\mathrm{i},\ext}$,
amplitude $-gJ_{\ext}$) inputs as
\begin{eqnarray}
r_{\mathrm{e},\ext} & = & r_{\mathrm{e},0}+r_{\mathrm{bal}}\quad r_{\mathrm{i},\ext}=r_{\mathrm{bal}}/g\label{eq:external_adjust}\\
\text{with }r_{\mathrm{e},0} & = & \frac{\mu_{i}-\mu_{\mathrm{loc.}}}{J_{\ext}\taum}\quad\text{and }r_{\mathrm{bal}}=\frac{\sigma_{i}^{2}-\sigma_{\mathrm{loc.}}^{2}-\taum r_{\mathrm{e,0}}J_{\ext}^{2}}{\taum J_{\ext}^{2}(1+g^{2})},\nonumber 
\end{eqnarray}
where $\mu_{\mathrm{loc.}}=\taum rKJ(1-\gamma g)$ and $\sigma_{\mathrm{loc.}}^{2}=\taum rKJ^{2}(1+\gamma g^{2})$
are the mean and variance due to local input from other neurons of
the network firing with rate $r$. From $L=\mathrm{const.}$ follows
that also $Kw=\mathrm{const.}$, so that \eqref{eq:corrtrans_theory}
predicts a scaling of the magnitude of the covariance functions in
proportion to $1/N$. Other network parameters are $d=3\ms$ and $g=5$.
The firing rate in the network is $r=23.6\Hz$.

In \prettyref{fig:phase_diagram} and \prettyref{fig:corr_components}
we use $N=10^{4}$, $g=6$, $J=0.1\mV$, and the delay $d$ as described
in the captions, the remaining parameters are as in \prettyref{fig:scaling_corr_functions}.

In \prettyref{fig:Okun_Lampl} we use $N=10^{4}$, $J=0.1\mV$, $g=5$,
and $d=2\ms$ and the remaining parameters as in \prettyref{fig:scaling_corr_functions}.
We obtain the filtered synaptic currents by filtering the spike trains
with an exponential filter of time constant $\taus=2\ms$. This results
in an effective filter for the cross covariances of \foreignlanguage{english}{$q(t)=\frac{\taum^{2}}{2\taus}e^{-|t|/\taus}$}.
The different contributions shown are $c_{I_{\Ex}I_{\Ex}}=q\ast(J^{2}pKa_{\Ex}+J^{2}K^{2}c_{\Ex\Ex})$,
$c_{I_{\In}I_{\In}}=q\ast(J^{2}pKa_{\In}+J^{2}K^{2}c_{\In\In})$,
$c_{I_{\Ex}I_{\In}}=-q\ast(gJ{}^{2}\gamma K{}^{2}c_{\Ex\In})$, and
$c_{I_{\In}I_{\Ex}}=-q\ast(gJ{}^{2}\gamma K{}^{2}c_{\In\Ex})$, where
$\ast$ denotes the convolution.

For \prettyref{fig:response_kernel}, we simulate two populations
of $N=1000$ neurons each. Each neuron receives independent background
activity from Poisson processes and in addition input from a common
Poisson process with rate $r_{c}=25\Hz$ causing in population 1 a
positive synaptic amplitude of $J$ and for population 2 a negative
synaptic amplitude $J$ ($J$ is given on the x-axis). The synaptic
amplitude of the background inputs is $J_{\Ex}=0.1\mV$ for an excitatory
impulse and $J_{\In}=-0.5\mV$ for an inhibitory impulse. The rates
of the excitatory and inhibitory background inputs are chosen so that
the first and second moments $\mu_{i}=\taum(J_{\Ex}r_{\Ex}+J_{\In}r_{\In}+J\, r_{c})=15\mV$
and $\sigma_{i}^{2}=\taum(J_{\Ex}^{2}r_{\Ex}+J_{\In}^{2}r_{\In}+J\, r_{c}^{2})=10\mV$
are independent of $J$. The spikes produced by each population are
triggered to the arrival of an impulse in the common input and averaged
over a duration of $10\sec$ to obtain the impulse response.

\end{document}